\documentclass[12pt,a4paper]{article}

\usepackage{amsmath,amssymb,amsfonts}
\usepackage[dvips]{graphicx}
\usepackage{epsfig}

\usepackage{cite}
\usepackage{amsmath,amssymb,amsfonts,graphicx}
\usepackage{epsfig}
\usepackage[left=3.2cm,top=3.6cm,right=3.2cm,bottom=3.6cm,bindingoffset=0cm]{geometry}

\newcommand{\beq}{\begin{eqnarray}}
\newcommand{\eeq}{\end{eqnarray}}
\newcommand{\bea}{\begin{eqnarray}}
\newcommand{\eea}{\end{eqnarray}}

\newcommand{\bec}{\begin{center}}
\newcommand{\eec}{\end{center}}

\def\x{\tilde{x}}
\def\z{\tilde{z}}

\numberwithin{equation}{section}

\begin{document}

\title{
\vskip 10pt
\bf{ Instanton Bags, High Density  Holographic QCD  and Chiral Symmetry Restoration}
\vskip 20pt}
\author{S. Bolognesi\\[20pt]
{\em \normalsize
Department of Physics “E. Fermi” and INFN, University of Pisa}\\[0pt]
{\em \normalsize
Largo Pontecorvo, 3, Ed. C, 56127 Pisa, Italy}\\[3pt]
{\em \normalsize   and }\\[3pt]
{\em \normalsize
Department of Mathematical Sciences,}\\[0pt]
{\em \normalsize Durham University, Durham DH1 3LE, U.K.}\\ [5pt]
{\normalsize Email: stefanobolo@gmail.com}
}
\vskip 10pt
\date{October 2014}
\maketitle
\vskip 0pt
\begin{abstract}

We describe the simplest example of an instanton bag in Euclidean space. It consists of a monopole wall and a Kaluza-Klein monopole wall, lifted to one higher dimension,  trapping the instanton charge in the middle.  This object has finite instanton density in a three-dimensional volume.

Baryon physics in holographic QCD models gets translated into a multi-instanton problem in the bulk, and a state with a high density baryonic charge consists of a non-diluted multi-instanton solution. The instanton bag is a good candidate for this high-density state.  We compute its parameters via moduli stabilization.  Chiral symmetry restoration is exhibited by this state, and it is a direct consequence of its non-diluted features.

\end{abstract}
\newpage

\section{Introduction}

Holographic QCD (HQCD) has been the subject of intense studies because  it provides an environment  to address strong coupling QCD questions with the use of semi-classical computations.   Baryons of the  QCD-like theory defined on the boundary correspond to instantons in the bulk.   Thus, having a finite baryonic density in the dual theory  corresponds  to having a finite density of instantons in the bulk.  The problem of the high-density phase of  QCD is thus translated into a multi-instanton problem, which, even at the classical level, remains quite a  difficult challenge, both numerically and analytically.
Recently with P.~Sutcliffe, we considered a toy model in $2+1$ dimension in which the high-density phase can be studied both analytically and numerically \cite{Bolognesi:2013jba}.  This model showed that the dilute approximation breaks down, as expected,  exactly  at the  densities  when the solitons are sufficiently close to start to populate the holographic direction.

Solitons are not hard objects but can compenetrate with one another.  A high-density phase of solitons may be in the form of a collective structure,  which is completely different from a  coarse-grained version of a large number of small constituents.  This phenomenon has been observed both for vortices and monopoles \cite{miei} and is known as `solitonic bag'.   We still do not have a proper  soliton bag description  for multi-instanton configurations, although it has been suggested that this should somehow exist \cite{Sutcliffe:2012pu,Harland:2012cj}.  One obstacle in finding such a solution for instantons is that for the Yang-Mills theory in Euclidean space, instantons do not have an intrinsic scale.    Thus, it is not uniquely defined how to make a  high-density limit of instantons;  the sizes must be specified in  the limiting process.   HQCD is a situation in which  instantons do have an intrinsic  scale.   This scale is fixed by a balance between the curvature of the background geometry and the Chern-Simons coupling, which makes the instanton electrically charged.  We may expect that a large number of instantons in HQCD could form  a bag structure at large enough density.

QCD at high-density is expected to exhibit a phase transition in which  chiral symmetry is restored.   Moreover, there is a competition between two other possible instabilities:  color superconductivity and chiral density waves.   The second one is favored in the large $N_c$ limit \cite{Deryagin:1992rw,Shuster:1999tn}.
Holographic models of QCD, being particular cases of large $N_c$  QCD-type models,  are thus expected to have chiral symmetry restoration together with chiral waves.
Recently, these questions have been addressed in HQCD \cite{deBoer:2012ij}, where it has been found that, using an instanton fluid  approximation, there is no sign of those phases.  We will show that they are instead realized in the instanton bag background.
Another generic expectation from large $N_c$ QCD is the existence of a `quarkyonic phase' \cite{McLerran:2007qj}. This is generally believed to correspond, in HQCD, to the instantons populating the holographic direction \cite{Kaplunovsky:2012gb,deBoer:2012ij,Kaplunovsky:2013iza}, so that many instantons share the same $3$D spatial section.
Yet a different phase in HQCD has been discussed in  \cite{Rho:2009ym,Ma:2013ooa} and named `dyonic salt'. This phase is analogue to the half-Skyrmion phase in the high-density Skyrme model and, for this reason, it has been argued to exhibit chiral symmetry restoration.   The instanton bag we discuss in the present paper can be considered as an extension of the dyonic salt phase to densities where the instantons start to populate the holographic direction. In a sense, this paper is a reconciliation of various approaches. Our state is also similar to the almost homogeneous state considered in \cite{Rozali:2007rx}, but with a different motivated ansatz.

The paper is organized as follows. In Section \ref{instbag} we describe an instanton bag solution in flat space. In Section \ref{holoqcd} we review some generic features of HQCD at finite density. In Section \ref{instbagembedding} we embed the instanton bag in HQCD and study the moduli stabilization.  In Section \ref{string} we review the top-down derivation of the Sakai-Sugimoto model.   In Section \ref{chisymrest} we discuss the phenomenon of chiral symmetry restoration at large densities. We conclude in Section \ref{concl}.

\section{An instanton bag from monopole walls}
\label{instbag}

We consider a Yang-Mills (YM) theory in $5$D Minkowski  space-time with gauge group $SU(2)$. The  action is
\beq
\label{ym5}
S_{YM5} = - \int dt d^4 x \frac{1}{4g^2} F_{\mu \nu}^a F^{\mu\nu a} \ .
\eeq
Instantons correspond to particles with mass $8 \pi^2 /g^2$.
We will consider solutions that are periodic in one direction, and thus we can formally compactify the $x_3$ direction on a circle of radius $R_3$ (we chose the direction $3$ for later convenience).
The sector of configurations that are also invariant along $x_3$ have the 4D Yang-Mills-Higgs (YMH) action
\beq
\label{m4}
S_{YMH4} = - \int dt dx_1 dx_2 dx_4 \frac{ \pi R_3}{2 g^2} \left( F_{\mu \nu}^a F^{\mu\nu a}  + D_{\mu} \phi^a D^{\mu} \phi^a\right) \ ,
\eeq
where $\phi$ is just another name for $A_3$.  This is the first term in a Kaluza-Klein (KK) series expansion.
The monopole wall studied in \cite{Lee:1998isa,Ward:2006wt} (see also \cite{Bolognesi:2010nb,Cherkis:2012qs,Sutcliffe:2011sr,Hamanaka:2013lza,Cherkis:2014vfa,Maldonado:2014gua}) is a solution of the
Bogomoln'y equations $F_{ij}= \epsilon_{ijk}D_k\phi$ for the YMH action (\ref{ym5}) and  thus, lifted to $4$D by keeping  the fields $x_3$ independent, it is also a self-dual instanton solution satisfying $F = \widetilde{F}$.

\begin{figure}[h!t]
\epsfxsize=14.0cm
\centerline{\epsfbox{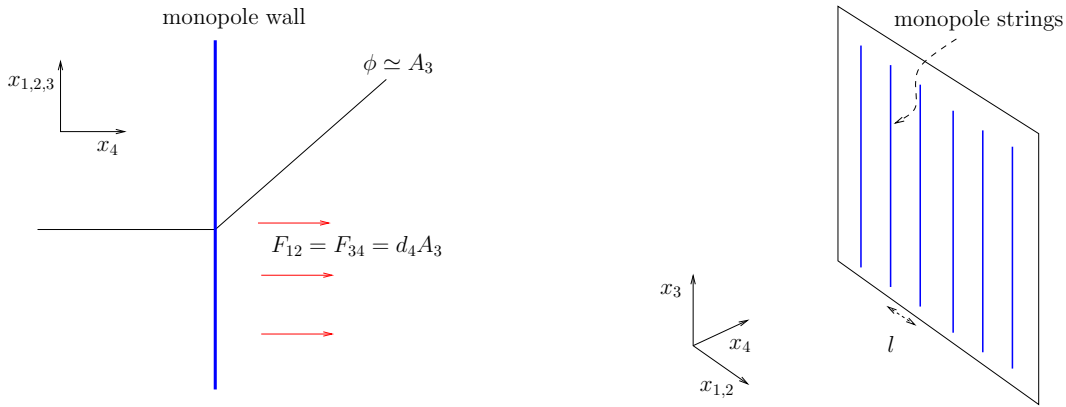}}
\caption{{\footnotesize Monopole wall}}
\label{monopolewall}
\end{figure}

A sketch of the monopole wall solution is given in Figure \ref{monopolewall}. The wall is located at a fixed position in $x_4$ and is extended in the $x_{1,2,3}$ directions.  The  wall separates two phases, one on the left in which $A_3$ is constant and the field strength is vanishing, and the other  on the right in which $A_3$ grows linearly with $x_4$  and the magnetic $B$ field is constant. This is in the singular gauge in which the fields far from the wall are directed in a fixed direction in the su$(2)$ algebra, and Dirac string singularities are on the empty left side of the wall.  The Bogomoln'y equation  $B=F_{34}$ implies that  $A_3$ is growing linearly as a function of $x_4$.  Details about the non-Abelian nature and its lattice structure are all contained in a small strip near the wall and can be neglected at large distances.
When this solution is lifted to $4$D, the monopoles become monopole strings extended in the $x_3$ direction and the wall becomes a three-dimensional object.

The monopole wall solution is specified  by the following set of parameters. The first is the  area of the unit monopole lattice  in the $1$-$2$ plane.  We name $l$  the typical distance between the constituents monopoles and $l^2$ is the area of the unit lattice in the $1$-$2$ plane.  The magnetic field sourced by the wall is then related to the microscopic structure by
\beq
\label{brelation}
F_{12} = B \, t_{{\rm su}(2) } \qquad \qquad B = \frac{4\pi}{ l^2} \ ,
\eeq
which is obtained by requiring that one unit of magnetic flux passes through each cell.
Also, $t_{{\rm su}(2)}$ is any  normalized generator of $SU(2)$, for example $t_3 = {\rm diag}(1/2,-1/2)$.  In this singular gauge, the field $F_{12}$ it is  constant far from the wall.
The field $A_3$ is then given by
\beq
\label{adjoint}
\left.A_3 \right|_{x_3 <0} = 0 \qquad \quad {\rm and}  \qquad \quad
\left.A_3 \right|_{x_3 >0} = B x_3 \,  t_{{\rm su}}(2) \ .
\eeq
The transverse thickness of the wall is of order
\beq
\label{thicknesswall}
\delta \simeq \frac{1}{ \sqrt{B}} \ ,
\eeq
being the scale where massive $W$ bosons start to condense near the wall.
As long as we are interested in length scales much greater than the lattice size $l$ and the transverse thickness $\delta$, the microscopic details of the monopole wall are not relevant.
The monopole wall solution has a constant instanton charge and also constant  energy density on the right side and zero on the left side.

The monopole wall, at the microscopic level, necessarily breaks translational invariance. This is somehow analogue to the non-existence of exactly spherical symmetry monopoles for charge higher than one \cite{Weinberg:1976eq}. The lattice structure of the monopole wall has been studied in some detail using numerical techniques in \cite{Ward:2006wt} for the BPS case and in \cite{Sutcliffe:2011sr} for AdS$_4$ where the hexagonal lattice has been found to be slightly favourite energetically. For the present paper, we do not  need to know the details of  the lattice structure because we consider only the large scale properties.

A monopole wall in isolation is not enough to have finite instanton density in ${\mathbb R}^3$.  For this, we also need a wall of Kaluza-Klein  monopoles, or a `KK monopole wall'.  KK monopoles are  solutions of YM equations  on $\mathbb{R}^3 \times S^1$ with a non-trivial dependence on the compactified direction. They carry the same instanton charge of the monopole but an opposite magnetic charge. One instanton on $\mathbb{R}^3 \times S^1$ is then  decomposable  into one monopole plus  one  KK monopole \cite{Lee:1997vp,Lee:1998bb,Kraan:1998pm}, and the distance between the constituents is continuously connected with the scale modulus of the instanton.   We will  show that with a  monopole wall and a KK monopole wall, we can create a configuration that has finite instanton density in a three-dimensional volume.

When the theory is  compactified on a circle  $S^1$, so with the periodic identification $ x_3 \equiv x_3 + 2 \pi R_3 $, we can make a gauge transformation that is topologically distinct from identity map
\beq
\label{gauge}
U(x_3) = e^{-i x_3 t_{{\rm su}(2)} / R_3} \ .
\eeq
This function  goes from plus to minus the identity  as $x_3$ completes its period.
If the theory contains only adjoint fields, the gauge transformation (\ref{gauge}) connecting the two elements of the centre of the group $1$ and $-1$ is thus single valued.  This gauge transformation shifts the gauge field by a constant
\beq
\label{large}
A_3 \ \ \longrightarrow \ \ A_3 - \frac{t_{{\rm su}(2)}}{R_3} \ .
\eeq
Therefore,  the gauge field $A_3$ assumes value in a T-dual circle.

The KK monopole is an ordinary monopole transformed by a large-gauge transformation action plus a  global gauge transformation that flips the sign of the generator $t_{{\rm su}(2)} \to - t_{{\rm su}(2)}$.
The KK monopole has opposite sign relation between the instanton charge and the monopole charge. For example if we normalize so that the monopole has $(n_{mon},n_{inst})$  charge equal to $(1/2,1/2)$,
the KK monopole has charges $(-1/2,1/2)$. 
So the monopole together with a KK monopole has exactly the charge of an instanton $(0,1)$.
The asymptotic value of the Higgs field is $\phi \equiv A_3 \simeq t_{{\rm su}(2)}/2 R_3$, so the large-gauge transformation (\ref{large}) plus the global transformation leave it invariant, and the monopole and KK monopole can be glued together. The global transformation flips the sign of both $d_{r} A_3$ and the magnetic field so that it does not affect the instanton charge.

\begin{figure}[h!t]
\epsfxsize=18.0cm
\centerline{\epsfbox{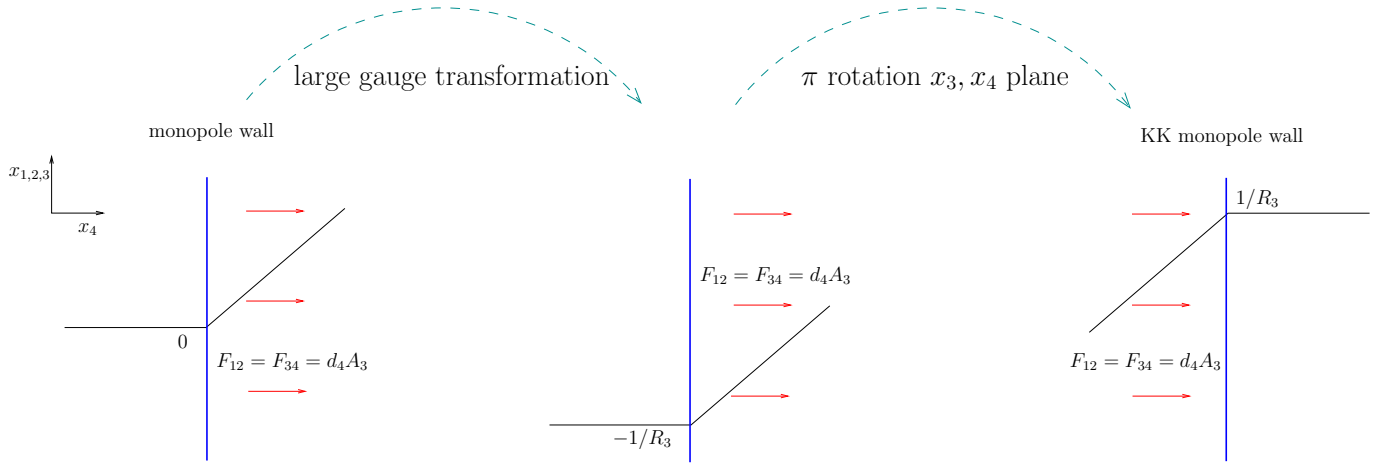}}
\caption{{\footnotesize Sequence of transformations from monopole wall to KK monopole wall, first the large gauge transformation and then the $\pi$ rotation in the $x_3,x_4$ plane.}}
\label{transformations}
\end{figure}

The KK monopole wall can be obtained in a similar fashion by applying  two transformations to the monopole wall.
First we use  the same large-gauge  transformation  (\ref{gauge}) and then  a $\pi$ rotation in the $x_3,x_4$ plane:  $x_{3,4} \to -x_{3,4}$ and  $x_{1,2} \to x_{1,2}$.  Gauge fields are vectors, so $A_3$ flips sign with this rotation.
On the initial left side of the wall, the empty half $x_4<0$,   the composition of the two transformations gives  the following:
\beq
A_3 =  0 \ \  \longrightarrow \ \  - \frac{t_{{\rm su}(2)}}{R_3} \ \  \longrightarrow  \ \  + \frac{t_{{\rm su}(2)}}{R_3} \ .
\eeq
Moreover, the empty side is moved to the right side of the wall $x_4>0$.
On the  non-empty side, the one filled with magnetic field $x_4>0$,  the  field  $F_{12}$ remains unchanged because $x_{1,2}$ are not affected by the rotation. Also $d_4 A_3$ is unchanged because both $d_4$ and $A_3$ flip the sign.  Therefore, both monopole wall and KK monopole wall satisfy the BPS equation with the same sign. The transformation we have described is exactly what we need to glue together the two walls.  The two  transformations are shown  in Figure \ref{transformations}.

\begin{figure}[h!t]
\epsfxsize=11.0cm
\centerline{\epsfbox{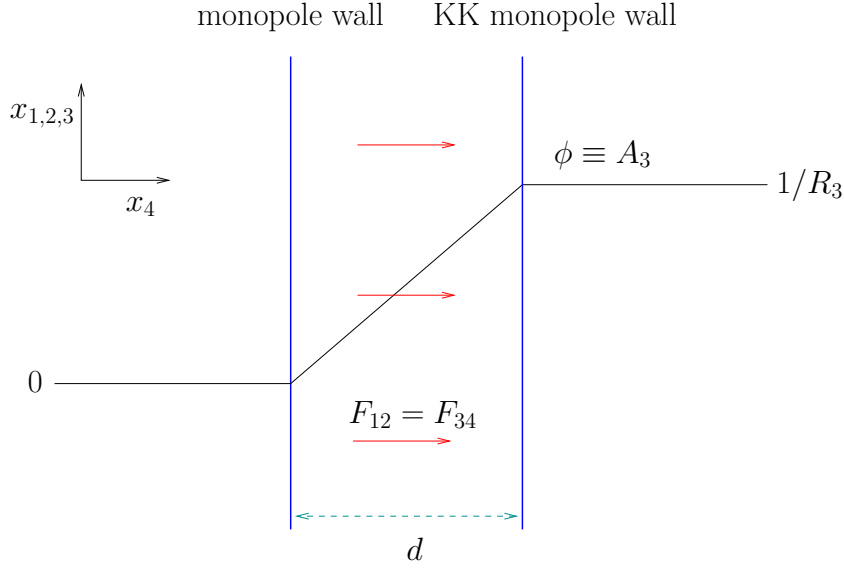}}
\caption{{\footnotesize Monopole wall and KK monopole wall with instanton charge inside the two plates.}}
\label{instantonbag}
\end{figure}

 A  monopole wall together with a KK monopole wall is capable of trapping the instanton charge in the middle of the two plates in the $x_4$ direction.  They can be glued together since  the $B$ field and the derivative of $A_3$ have the same sign (see Figure \ref{instantonbag}).
 The microscopic scales are the lattice size $l$ and the transverse thickness $\delta$. The distance between the two walls is  $d$.  $A_3$ grows linearly between the two walls from $0$ to $1/R_3$. The  BPS equation gives the following relation:
\beq
\label{bogorelation}
B = \frac{1}{d R_3} \ .
\eeq

The instanton charge  is still infinite in the directions $x_{1,2,3}$, but  it is finite if interpreted as a density in the unit of three-dimensional volume. If we call $Q$ the instanton density in a three-volume $dx_1 dx_2 dx_3$,  we have the relation
\beq
\label{irelation}
Q = \int dx_4 \frac{1}{32 \pi^2} F^a_{\mu\nu} \tilde F^{\mu\nu a} = \frac{B}{ 8 \pi^2 R_3} \ .
\eeq
We count the number of parameters of the solution.  In total there are four parameters, $l$, $B$, $R_3$, $d$, and two relations between them, (\ref{brelation}), (\ref{bogorelation}). Moreover, we want the  instanton charge (\ref{irelation}) to be fixed.  We thus  remain with a one-parameter family of solutions, which we can take to be the separation $d$.   Rewriting everything as a function of the distance $d$, the coupling $g$, and the instanton charge density $Q$,   we have the following set of relations:
\beq
\label{parametersflat}
\delta \simeq l=\frac{2 \sqrt{\pi}}{\sqrt{B} } \quad  \qquad R_3=\frac{1}{2 \pi \sqrt{2 Q d}} \quad \qquad  B=2 \pi \sqrt{\frac{2 Q}{d}} \ .
\eeq

The instanton bag we just discussed is very similar to the composite monopole walls studied in \cite{Cherkis:2012qs,Hamanaka:2013lza,Cherkis:2014vfa,Maldonado:2014gua}.  These solutions have an hyper-Kahler moduli space which, for two walls only, is $4+4$-dimensional. The relative $4$ moduli are  distance $d$.  a $U(1)$ relative phase, and a shift in the lattice positions of the two walls.  The distance $d$ is, for the present paper, the only modulus of these four that  we will consider because the others affect only the microscopic structure.

For the bag approximation to be valid,  we want to be in a regime in which we can neglect the microscopic structure of the wall,  and thus we want the distance $d$ to be much greater than  $l$ and $\delta$.  The most stringent condition of the two's is
\beq
\label{microstructure}
 d \gg l \qquad \quad  \Longrightarrow \qquad \quad \sqrt[4]{d^3 Q} \gg 1 \ .
\eeq
Note that, even by keeping  fixed $Q$ and $g$,  we can always reach a point in the moduli space where $d$ is large enough to make the approximation valid.  Yet another way to satisfy the condition is to  increase $Q$ while keeping $d$ and $g$ fixed.

It may be instructive to avoid the direct use of the Bogomoln'y equation (\ref{bogorelation}) and use instead the minimization of the energy.  The energy density per unit of volume $dx_1  dx_2 dx_3$ is
\beq
E =  \frac{1}{2g^2}   \int_{-d/2}^{d/2}  dx_4  \left(  {F_{12}^a}^2  +  {F_{3 4}^a}^2 \right) \ ,
\eeq
which then becomes
\beq
E =  \frac{1}{2g^2} \left(  d B^2  +  \frac{64 \pi^4 Q^2}{d B^2} \right)  \ .
\eeq
Minimizing by keeping $Q$ fixed, we determine the value of $d B^2$ and the corresponding energy density
\beq
d B^2 = 8 \pi^2 Q \qquad \qquad E = \frac{ 8 \pi^2 Q}{ g^2}\ \ ,
\eeq  
where (\ref{bogorelation}) is recovered, and one parameter between  $d$ and $B$ disappears from $E$ and remains thus a free modulus.  This minimization strategy is the same we will use in Section \ref{instbagembedding} in the HQCD context, where the BPS equation is no longer valid.

\begin{figure}[h!t]
\epsfxsize=14.5cm
\centerline{\epsfbox{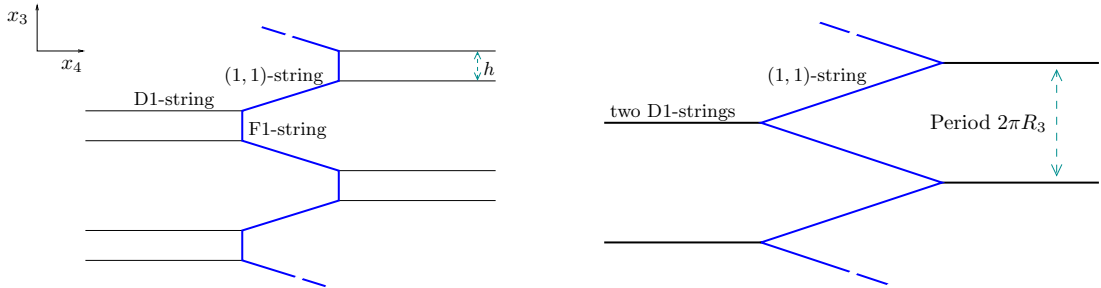}}
\caption{{\footnotesize Web of string junctions in the $x_{3,4}$ plane. The basic constituent is the three string junction D1-F1-$(1,1)$.  The structure is periodic in the $x_{3}$ direction. The second is the web for the limit of coincident D1-strings. This configuration corresponds to the instanton bag of Figure \ref{instantonbag}.}}
\label{braneint}
\end{figure}

A brane construction in string theory  provides a very intuitive realization  of the monopole wall \cite{kim,Harland:2012cj}.
Monopoles are realized as D1-strings stretched between two D3-branes in type IIB string theory.  We may take the D3 worldvolume to be stretched along the directions $x_{0,1,2,4}$ and the D1-string along $x_{0,3}$ with the two D3-branes separated in the $x_3$ direction.  A monopole wall, in its simplest lattice realization,  is an ${\mathbb R}^2$ periodic configuration, which we take along $x_{1,2}$.  A series of transformation brings this configuration  to a simpler system.  First it can be turned into a D3-F1 system by an  S-duality.   Then, because we take a periodic configuration  along $x_{12}$, we can compactify on a torus and perform T-duality on both directions.  After these transformations, the D3-brane becomes a D1-string along the directions $x_{0,4}$, and the D1-string becomes a fundamental F1-string along $x_{0,3}$. The monopole wall is thus transformed into a web of connected D1 and F1 strings periodic in $x_3$.   The basic building block of this web is a string junction among a D1, an F1, and a third dyonic $(1,1)$-string  with a certain angle in the  $x_{3,4}$ plane dictated by the balance of the tensions.  
The instanton bag configuration  is then described by the web in Figure \ref{braneint}. The junctions have to be placed in a periodic configuration in $x_3$ with the chain resulting from the KK monopole wall shifted of half-period respect to the chain coming from the monopole wall.  To obtain the instanton bag we take this periodic web of string junctions, and perform all the previous dualities in reverse order. When we arrive at the original D3-D1 configuration, we perform another T-duality in the direction $x_3$ and we get the D4-D0 system.

As  it is clear from the string web illustration in Figure \ref{braneint}, the monopole wall does not necessarily have to have a zero Higgs field on its empty side. The value of the Higgs field is in fact another parameter, which we call $h$ (the existence of this extra freedom was observed in \cite{Cherkis:2012qs}). For $h \neq 0$, the monopole wall has an intrinsic tension given by
\beq
T = \frac{B h}{g^2}  = \frac{M_{mon}}{l^2} \ ,
\eeq
which is consistent with the mass of a single BPS monopole with an asymptotic Higgs field equal to $h$
\beq
M_{mon} = \frac{4 \pi h}{g^2} \ .
\eeq
The two configurations in  Figure \ref{braneint} have exactly the same energy.  What is gained from the monopoles mass is lost from the fact that the two walls are closer.
In HQCD, this is no longer a free modulus, but must enter the minimization procedure like $d$ and $B$. We will see that the $h=0$ is, in general, the energetically favourite state.

We conclude this section with a discussion of the gauge in which we want to prepare our system.
When we embed the monopole wall  and KK monopole wall pair in holography, we want to choose a gauge that is the most convenient for the AdS/CFT dictionary. Usually, this is the gauge in which the gauge field is zero at the UV boundary, which are the two asymptotic limits $x_4 \to \pm \infty$.
\footnote{If the gauge field is not zero at the boundary but gauge equivalent to zero, and if we do not want to interpret it as a source, the AdS/CFT dictionary also must be modified accordingly.}
The gauge we used for the previous discussion is not of this kind because  there are  Dirac strings singularities, one for every period of the monopole wall and KK monopole wall.
To eliminate the Dirac strings we can go to the analogue of the hedgehog gauge for the 't Hooft-Polyakov monopole. In the middle of the two walls the gauge transformation is a function $U(x_1,x_2)$ so that the adjoint field $U(x_1,x_2) t_{{\rm su}(2)} U(x_1,x_2)^{-1}$ winds around $SU(2)/U(1)$ once for every period of the wall lattice.
This is particularly relevant for later application in HQCD. The magnetic fields $F_{12}$ and $F_{34}$ fluctuate around the su$(2)$ algebra so that every wave coming from the empty side of the wall in a fixed su$(2)$ generator (which in HQCD are dual to the vector meson states in a given isospin state) interacts with some magnetic field.
This gauge is convenient to check that the instanton charge is equal to the flux of the Chern-Simons current
\beq
K^{\mu} = \frac{1}{16 \pi^2}\epsilon^{\mu\nu\rho\sigma} \left( A^a_{\nu} F^a_{\rho,\sigma} + \frac{2}{3} \epsilon_{abc}A^a_{\nu} A^b_{\rho} A^c_{\sigma}\right) \ ,
\eeq
\begin{figure}[h!t]
\epsfxsize=10cm
\centerline{\epsfbox{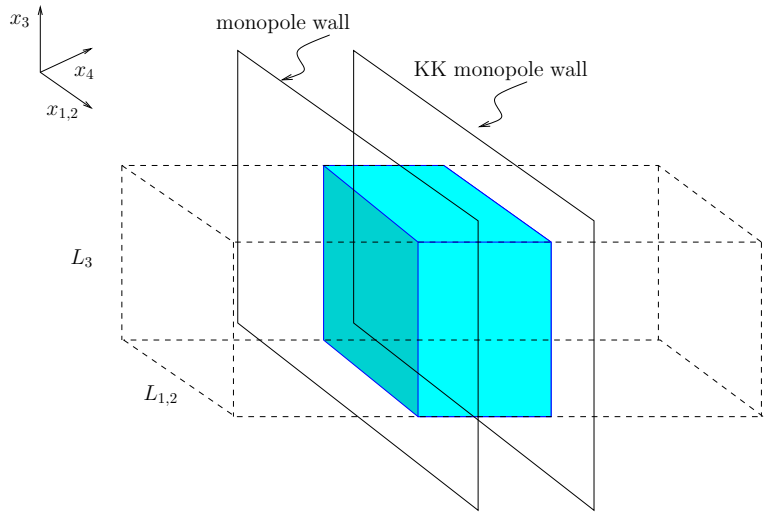}}
\caption{{\footnotesize Check of the Chern-Simons flux.}}
\label{fluxcs}
\end{figure}
We take a $4$D volume as in Figure \ref{fluxcs} that encloses part of the instanton bag.
The gauge field vanishes at the left side of the monopole wall and at the right side of the KK monopole wall. So the Chern-Simons flux can escape only through the volumes $1$-$3$-$4$, $2$-$3$-$4$, and $1$-$2$-$4$ inside the two walls.  No term from $\epsilon_{abc}$ in the Chern-Simons currect gives any contribution. Moreover there is no contribution from the $1$-$2$-$4$ volume. We just have to compute the flux from $1$-$3$-$4$ and  $2$-$3$-$4$. The total instanton charge $I$ is then
\beq
I = \int dx_{i} dx_3 dx_4  \frac{1}{8 \pi^2}  A^a_{i} F^a_{34}  = \frac{B   }{8 \pi^2 R_3}  L_1 L_2 L_3 \ ,
\eeq
where $i=1,2$. This is consistent with the instanton density (\ref{irelation}).

\section{HQCD and  finite baryon density}
\label{holoqcd}

The Sakai-Sugimoto model, at low energy, consists of a $U(2)$ gauge theory in the following $5$D  warped metric background:
\beq
\label{metric}
ds^2 = H(z) dx_{\mu} dx^{\mu} +  H(z)^{-1} dz^2 \qquad \qquad
H(z) = \left(1+\frac{z^2}{z_0^2}\right)^{2/3} \ .
\eeq
The length scale $z_0$ is related  to the inverse of the dynamical scale in the dual QCD-like theory.  From now on we set $z_0=1$.
The action is a sum of the Yang-Mills term plus a $U(2)$ Chern-Simons term:
\bea
\label{actionymcs}
S = -\frac{N_c \lambda}{216 \pi^3}   \int d^4 x dz \sqrt{-g} \  \frac{1}{2} {\rm tr} \left({\cal F}_{\Gamma\Delta} {\cal F}^{\Gamma\Delta}\right) +  \frac{N_c}{24 \pi^2} \int d^4 x dz \ \omega_5({\cal A}) \ .
\eea
 We  use the indices conventions as follows:  $\Gamma,\Delta,... = 0,1,2,3,z$;  $  \mu,\nu,... = 0,1,2,3$;  $I,J=1,2,3,4$  and $i,j,...=1,2,3$.
The factors $N_c$ and $\lambda$ are respectively the number of colours and the 't Hooft coupling of the dual theory.
We can decompose the gauge field into non-Abelian and Abelian components $U(2) = SU(2) \times  U(1)$:
\beq
{\cal A}_{\Gamma} = A_{\Gamma} + \frac{1}{2} \hat{A}_{\Gamma} \qquad  \qquad
{\cal F}_{\Gamma\Delta} = F_{\Gamma\Delta} + \frac{1}{2} \hat{F}_{\Gamma\Delta} \ .
\eeq
In the following, we are only interested only in static configurations, so we restrict to the  following ansatz:
\beq
A_I,\hat{A}_0=A_I(x_{I}),\hat{A}_0(x_{I}) \qquad \qquad  A_0,\hat{A}_{I} = 0,0 \ .
\eeq
The action is then reduced to
\bea
\label{staticaction}
{\cal S}  &=& \int d^4 x dz \left\{   \frac{1}{2 H^{1/2}} (\partial_i \hat{A}_0)^2 + \frac{H^{3/2}}{2} (\partial_z \hat{A}_0)^2  - \frac{1}{2 H^{1/2}} {\rm tr}  \left( { F}_{ij}^2 \right) - H^{3/2} {\rm tr}  \left( {F}_{i  z}^2 \right) \right\}     \nonumber \\
&&  +  \frac{1}{\Lambda} \int d^4 x dz \ \hat{A}_0 \ {\rm tr}\,( {F}_{IJ}{ F}_{KS}) \,  \epsilon^{IJKS} \ ,
\eea
where we have rescaled for convenience the action and the 't Hooft coupling as
\beq
\label{rescalingsl}
  {\cal S} = \frac{ 64 \pi^2 }{\Lambda N_c}  S \quad \qquad {\rm and} \qquad \quad \Lambda = \frac{8 \lambda}{27 \pi} \ .
\eeq

For large 't Hooft coupling,  the Chern-Simons term is parametrically suppressed with respect to the Yang-Mills term.  The instanton in the large $\Lambda$ limit can thus be approximated by a BPS ansatz \cite{Hong:2007kx,Hata:2007mb}
\beq
\label{radialansatz}
A_{I} = - {\sigma}_{IJ} x_{J} \,  \frac{1}{\rho^2 + l^2}
\eeq
with $\sigma_{ij} = \epsilon_{ijk}\sigma_k$ and $\sigma_{iz} = - \sigma_{zi} =  \sigma_i$.
The size of the instanton $l$ is then obtained by minimizing the energy restricted to the BPS moduli space. This gives
\beq
l = \frac{3^{1/4} \, 2^{7/4}}{  5^{1/4}  \Lambda^{1/2}  }\qquad {\rm and} \qquad \quad
{\cal E} = 8 \pi^2  \left( 1 + \frac{2^{7/2} }{ 15^{1/2}\Lambda} + \dots \right) \ .
\eeq
At large $\Lambda$, the size of the instanton is much smaller that the curvature scale of the metric\footnote{Upon substitution $l = 3^{7/4}\pi^{1/2}2^{1/4}/5^{1/4} \lambda^{1/2}$ which is consistent with the results in the literature \cite{Hong:2007kx,Hata:2007mb,Bolognesi:2013nja2,Hong:2014bra}.}, which is why the BPS profile function provides a good approximation of the true solution, at least in the almost-flat region of the metric.  Moreover, the correction to the BPS mass is  a sub-leading term of order ${\cal O}\left(1/\Lambda\right)$. The solution flows exactly to the BPS for $\Lambda \to \infty$  but only in rescaled coordinates \cite{Bolognesi:2013nja2}, so the large-distance properties are not captured by the BPS ansatz.  Large distance properties are not relevant for high-density QCD due to the small distance between baryons will not discuss them in this paper.

At finite  densities, the instantons are distributed on a lattice configuration, and they begin to populate the $3$D space by all sitting at the bottom of the gravitational potential at $z=0$.  Their average distance is thus $d = Q^{-1/3}$, where $Q$ is the $3$D instanton density.   This configuration  remains valid as long as the typical distance $d$ is larger than the size of the single instantons, and thus for densities $Q  \ll \Lambda^{3/2}$.  In this regime,  the  energy density is dominated by the BPS term, and thus linear in $Q$.
In the ensemble in which we keep the chemical potential fixed, a phase transition to hadronic matter is expected to take place when the average distance is of order of the curvature scale, thus $Q \simeq 1$. This is when the attractive force due to the pion-pion tail is comparable with the Coulomb repulsive force \cite{Kim:2009sr}. Above this scale $Q \gg 1$, the attractive force becomes irrelevant with respect to the Coulomb repulsive force.
 The Coulomb interaction, for any couple of instantons  at distance $\tilde{d}$, is order $E_{\rm Coulomb} \propto  1 / \Lambda^2 \tilde{d}^2$ and this is valid for $l \ll \tilde{d} \ll 1$. The $1/ \tilde{d}^2$ dependence is because we are in $4+1$ dimensions and the  $1 / \Lambda^2$ dependence is because $1 / \Lambda$ is the analogue of the electric charge.
We can compute the energy density of the $3$D lattice as a sum of the mass of the single constituents plus the correction due to the Coulomb interaction:
\beq
\label{energydensitylattice}
{\cal E}  =   8 \pi^2 Q  \left(1  + {\cal O}\left( \frac{ Q}{\Lambda^2} \right) \right) \ .
\eeq
Note that, for the Coulomb term,  we summed the  $1/ \Lambda^2\tilde{d}^2$ term for every pair of constituents, taking as cutoff $\tilde{d} \simeq 1$, where the Coulomb law is certainly  modified by the metric curvature.   The Coulomb correction  becomes comparable with the BPS energy  at densities of order $Q  \simeq \Lambda^{2}$, which are much bigger than $\Lambda^{3/2}$.

When the instanton density becomes of order $Q \simeq \Lambda^{3/2}$,  two interesting things happen almost simultaneously.  First, the instanton lattice is no longer diluted because the typical inner distance becomes comparable with the instanton size $d \simeq l$.  Second, the instantons begin to climb the holographic direction as we are going to see with the following simple estimate.
We can estimate the force acting on each single instantons.
The first force is the gravitational one,  $F_{\rm Grav} \propto  - \, \Lambda z$, where $z$ is the linear displacement around the bottom of the gravitational potential.  Then there is a Coulomb force due to the surrounding  instantons in the $3$D lattice. This force, projected  in the $z$ component, is proportional to $z/\tilde{d}^4$ for any pair of instantons. The total force  is thus $F_{\rm Coulomb} \propto  + \, Q^{4/3} z $. This force is pulling the instantons away from the bottom of the gravitational potential because it is repulsive. The instantons begin to move away from the bottom of the gravitational potential when the two forces are comparable $F_{\rm Grav} \simeq F_{\rm Coulomb}$.  This condition is the same as the breaking of the diluted approximation.
The fact that these two changes both happen at the densities  $Q \simeq \Lambda^{3/2}$ is not a coincidence; it has also been discussed in \cite{Ghoroku:2012am} and observed in the toy model \cite{Bolognesi:2013jba}. The agents that  stabilize the instanton radius are, in fact,  the same that  decide  when the instanton lattice prefers to move the holographic direction.

\begin{figure}[h!t]
\epsfxsize=8.0cm
\centerline{\epsfbox{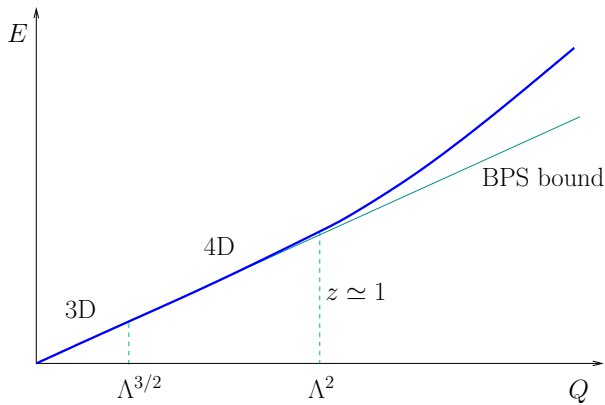}}
\caption{{\footnotesize Generic expectation for the energy density of the HQCD phases as function of instanton density $Q$.}}
\label{phases}
\end{figure}
At densities  $Q \simeq \Lambda^{3/2}$, the instantons start to fill the holographic direction as well. After that, another transition happens  at higher densities $Q \simeq \Lambda^2$.  Assuming the average distance remains of the order of the instanton scale  $d \simeq l$, the instanton $4$D lattice boundary reaches  the curvature scale $z \simeq 1$,  which is  when $Q$ is of order $\Lambda^2$.  This is also the density when the Coulomb energy becomes comparable with the BPS energy (\ref{energydensitylattice}), and thus the energy density strongly deviates from the BPS bound.  These phases are summarized in Figure \ref{phases}.

The information we extracted  so far is  based on qualitative estimates. To find the actual instanton solution for the high-density phase is a much harder task. A $4$D instanton lattice is specified by many parameters: the $3$D lattice strata, the relative orientation of the instantons in the internal space, the sizes of the instantons, and the depth of the various layers in the holographic direction $z$.  For densities $Q  \ll  \Lambda^2$, the problem simplifies since we can use a BPS ansatz and minimize the energy restricted to the BPS moduli space.
But even in this simplified case, the problem remains difficult to solve in full generality.
Some results  of the  self-dual configurations periodic on ${\mathbb T}^3$ and ${\mathbb T}^4$ have been obtained using a version of the ADHM transform \cite{Ford:2000zt}.  But the energy minimization would require  knowledge of the actual gauge fields and not just the ADHM data.

Our strategy in the rest of the paper is to use the instanton bag discussed in Section \ref{instbag} as an ansatz for the high-density phase. For the instanton bag approximation, we can compute the  fields and thus perform the energy minimization.  At the moment, we do not have sufficient control over other possible instanton configurations to compute their energy and thus decide which is the actual minimum for the high-density phase.

\section{Instanton bag embedded in the Sakai-Sugimoto model}
\label{instbagembedding}

We now embed the instanton bag studied in Section \ref{instbag}  into the Sakai-Sugimoto model.

We first use the BPS ansatz and find the best values of the moduli that  minimize the energy density at a fixed instanton density.  The BPS solution in flat space has two free moduli, the wall separation  $d$ and Higgs field $h$.  We now parametrize them with $\pm z_W$,  which are the positions of the monopole wall and KK-monopole wall in the $z$ direction,  and $\Delta$, which is the amount of instanton charge carried by the walls as a delta function.
The relation 
\beq
\label{bpsrelationqbzd}
Q = \frac{B^2 z_w }{4 \pi^2} + \Delta
\eeq
is valid for a self-dual solution 
and gives the instanton charge $Q$ as function of $z_w$, $B$ and $\Delta$.  We use it later to express $B$ as function of the other parameters.

The action (\ref{staticaction}) inside the two walls is reduced to
\beq
\frac{{\cal S}}{V_3} &=&  \int dt \int_{-z_w}^{z_w} dz  \left\{   \frac{H^{3/2}}{2} (\partial_z \hat{A}_0)^2  - \frac{ B^2 (1+ H^2) }{2 H^{1/2}  }  + \frac{4}{\Lambda}  \hat{A}_0 B^2 \right\} \ .
\eeq
From this, we find the equation for the Abelian electric potential $\hat{A}_0$:
\beq
\partial_z (H^{3/2}\partial_z \hat{A}_0) -  \frac{4}{\Lambda} B^2 = 0 \ .
\eeq
The solution for the electric field inside the walls is thus
\beq
\label{ina0}
\partial_z \hat{A}_0 = \frac{4 B^2 z}{ \Lambda (1+z^2)} \qquad \qquad \ {\rm for}\qquad |z| \leq z_w \ .
\eeq
The solution outside the walls is given by the solution of the equation without sources, matched with (\ref{ina0}) at $z=z_w$:
\beq
\label{outa0}
\partial_z \hat{A}_0 &=& \frac{16 \pi^2 Q}{\Lambda (1+z^2)} \qquad \,   {\rm for}\qquad |z| > z_w \ .
\eeq

The energy density is given by the sum of three contributions
\bea
{\cal E} =    \int_{0}^{z_w}  \left(   H^{3/2} (\partial_z \hat{A}_0)^2 + \frac{ B^2 (1+ H^2) }{ H^{1/2}  } \right) + 8 \pi^2 \Delta (1+z_w^2) +  \int_{z_w}^{\infty}  H^{3/2} (\partial_z \hat{A}_0)^2 \nonumber \\
\eea
with $A_0(z)$ given by (\ref{ina0}) and (\ref{outa0}).
We evaluate the integrals, and then expand in series of $z_w$, since the self-dual approximation is supposed to be valid only for $z_w \ll 1$.  The result is
\beq
&&{\cal E}(Q;z_w,\Delta) = 8 \pi^2 Q + \frac{128 \pi^5 Q^2}{\Lambda^2} +\nonumber \\ && \qquad \quad -\frac{256 \left(2 \pi ^4 Q^2+2 \pi ^4 Q \Delta -\pi ^4 \Delta ^2\right)}{3 \Lambda ^2} z_w  + \frac{8 \pi^2 (Q+8\Delta )}{9} z_w^2 + \dots \ .
\eeq
This energy density has to be minimized with respect to $z_w$ and $\Delta$  by keeping the instanton charge $Q$ fixed. The solution is given by
\beq
\label{minzwbps}
z_w = \frac{96 \pi^2 Q}{ \Lambda ^2} \qquad \qquad
\Delta = 0 \ .
\eeq
The energy density evaluated at the minimum is then
\beq
\label{bpsapproximation}
{\cal E}(Q) = 8 \pi^2 Q \left( 1 + \frac{16  \pi ^3 Q}{ \Lambda ^2} -\frac{1024 \pi ^4 Q^2}{\Lambda ^4} + \dots \right) \ .
\eeq
Note that the dominant term is the BPS bound, and the corrections are small if  $Q \ll \Lambda^2$.   This is also the regime in which $z_w \ll 1$.
Moreover, the bag approximation is valid when the wall microscopic structure is much smaller than $z_w$ (see (\ref{microstructure})) and thus, using (\ref{minzwbps}), becomes equivalent to  $Q \gg \Lambda^{3/2}$. Therefore, the BPS and bag approximations are both valid in the region of instanton densities $ \Lambda^{3/2} \ll Q \ll \Lambda^2$.

In the previous analysis we used the canonical ensemble:  we kept fixed the charge density $Q$ and minimized the energy density. 
We can also pass from the canonical to the grand-canonical ensemble.
The first way to recover the chemical potential with the usual thermodynamic relation
\beq
\label{muQfirstway}
\mu_Q = \frac{\partial {\cal E}(Q)}{\partial Q} =  8 \pi^2  + \frac{256  \pi ^3 Q}{ \Lambda ^2} - 24 * 1024 \frac{ \pi^6 Q^2}{ \Lambda ^4}  + \dots
\eeq
The second way, by using  the AdS/CFT prescription, is to compute the asymptotic value of $\hat{A}_0$. 
This can be done integrating the electric field in the two regions (\ref{ina0}) and (\ref{outa0}). The result is 
\beq
\hat{A}_0(\infty) = 
\hat{A}_0(0) + \frac{8 \pi^3 Q}{\Lambda} - \frac{768 \pi^4 Q^2}{\Lambda^3} + \dots
\eeq
where we have kept the first two orders on $Q/\Lambda^2$. To compare it with the chemical potential, we have first to find the proper normalization. The electromagnetic coupling comes from the Chern-Simon term of the action (\ref{staticaction})
\beq
{\cal S}_{CS}  = \frac{1}{\Lambda} \int d^4 x dz \ \hat{A}_0 \ {\rm tr}\,( {F}_{IJ}{ F}_{KS}) \,  \epsilon^{IJKS}  
= \int dt \frac{32 \pi^2}{\Lambda} \hat{A}_0 Q \ ,
\eeq
so the AdS/CFT prescription predicts the following relation between chemical potential 
$\mu_Q$ and the asymptotic value of $\hat{A}_0$:
\beq
\label{muQsecondway}
\mu_Q = \frac{32 \pi^2}{\Lambda} \hat{A}_0 (\infty)
\eeq
We can then see that the two ways to compute $\mu_Q$, (\ref{muQfirstway}) and (\ref{muQsecondway}), agree to every order if we choose $\hat{A}_0(0) = \frac{\Lambda}{4}$.

When $z_w$ becomes of order one, we can no longer use the self-dual approximation, although we can still use the instanton bag approximation. For this, we have to solve the profile function for $A_3(z)$ as it is generally modified by the metric curvature and is no longer linear in $z$.
The relation (\ref{bpsrelationqbzd}) is now given in more generality by the following:
\beq
\label{relationqbzd}
Q = \frac{B}{8 \pi^2} \int_{-z_w}^{z_w} \partial_z {A}_3 + \Delta \ .
\eeq

The action density between the two walls is
\bea
\label{staticactiondue}
\frac{{\cal S}}{V_3} &=& \int dt \int_{-z_w}^{z_w} dz \left\{  \frac{H^{3/2}}{2} (\partial_z \hat{A}_0)^2  - \frac{ B^2  }{2 H^{1/2}  } - \frac{  H^{3/2} }{2 } (\partial_z {A}_3)^2 + \frac{4}{\Lambda}  \hat{A}_0 B \partial_z {A}_3  \right\}  \ , \nonumber \\
\eea
from which we derive the equations for $\hat{A}_0$ and ${A}_3$:
\bea
\label{eqnbpsa0}
\partial_z \left(H^{3/2} \partial_z \hat{A}_0\right) -  \frac{4}{\Lambda}  B \partial_z {A}_3&=& 0\\
\partial_z \left(  H^{3/2}  \partial_z {A}_3 - \frac{4}{\Lambda}  \hat{A}_0 B  \right)&=& 0 \ .
\eea
The second one can be integrated
\beq
 \partial_z {A}_3 = \frac{ 4 \hat{A}_0 B - C\Lambda  }{\Lambda H^{3/2}} \ ,
\eeq
where $C$ is an arbitrary integration constant. The equation (\ref{eqnbpsa0}) , together with the boundary condition $ \partial_z \hat{A}_0(0)=0$, gives the solutions for $\hat{A}_0$
\beq
\label{a0ingen}
\hat{A}_0 =\hat{A}_0(0) + \frac{C \Lambda}{4 B} \left( 1- \cosh{\left(\frac{4 B \tan^{-1}{(z)}}{\Lambda}\right)}  \right) \ .
\eeq
The solution for $ \partial_z {A}_3$ is
\beq
\label{fz3}
 \partial_z {A}_3 = - \frac{C}{ H^{3/2}}\cosh{\left(\frac{4 B \tan^{-1}{(z)}}{\Lambda}\right)} \ .
\eeq
The relation (\ref{relationqbzd}) can now be written as
\beq
\label{relqcbzd}
Q = \frac{ C \Lambda }{16 \pi^2} \sinh{\left(\frac{4 B \tan^{-1}{(z_w)}}{\Lambda}\right)}   + \Delta
\eeq
Outside the walls, $A_3$ is constant while $\hat{A}_0$ is given by
\beq
\label{a0outgen}
\hat{A}_0 = c_1  -C \sinh{\left(\frac{4 B \tan^{-1}{(z_w)}}{\Lambda}\right)}\tan^{-1}{(z)} \ ,
\eeq
where $c_1$ is a constant that we do  not need for the moment.

The energy density is then given by
\bea
{\cal E} &=&   \int_{0}^{z_w}  \left(   H^{3/2}  (\partial_z \hat{A}_0)^2 + \frac{ B^2  }{ H^{1/2}  }  +  H^{3/2}  (\partial_z {A}_3)^2 \right) \nonumber \\
 && + \frac{\Delta (1+z_w^2)}{\pi^2} +  \int_{z_w}^{\infty}  H^{3/2} (\partial_z \hat{A}_0)^2 \ .
\eea
We can use the relation (\ref{relqcbzd}) to write the integration constant $C$ as function of the other parameters, and then the energy density becomes
\bea
\label{funcnores}
{\cal E}(Q;B,z_w,\Delta)  &=& \frac{32 \pi^4(Q-\Delta)^2  \sinh{\left(\frac{8 B \tan^{-1}{(z_w)}}{\Lambda}\right)}}{B \Lambda\left( \sinh{\left(\frac{4 B \tan^{-1}{(z_w)}}{\Lambda}\right)}\right)^2} +  B^2 \int_{0}^{z_w} \frac{1  }{ (1+z^2)^{1/3} }+  \nonumber \\
&&  + \frac{\Delta (1+z_w^2)}{\pi^2} +  \frac{256 \pi^4 (Q-\Delta)^2}{\Lambda^2} \left( \frac{\pi}{2} -\tan^{-1}{(z_w)} \right)
\eea
This is the function that has to be minimized with respect to the three parameters $B,z_w,\Delta$,  while keeping $Q$ fixed.
With the following convenient rescaling
\beq
B = b \Lambda \qquad  \qquad Q = q \Lambda^2 \ ,
\eeq
we can write the functional (\ref{funcnores}) as
\bea
\label{rescaled}
\frac{1}{ \Lambda^2 } {\cal E}(q;b,z_w) &=& \frac{32 \pi^4 q^2  \sinh{\left(8 b \tan^{-1}{(z_w)}\right)}}{b \left( \sinh{\left(4 b \tan^{-1}{(z_w)}\right)}\right)^2} +  b^2 \int_{0}^{z_w} \frac{1  }{ (1+z^2)^{1/3} }+  \nonumber \\
&&  +  256 \pi^4 q^2 \left( \frac{\pi}{2} -\tan^{-1}{(z_w)} \right) \ .
\eea
Numerical examples, for a given value of $q$, shows that the minimum always exists and is at $\Delta=0$ and for some finite value of $B,z_w$.

An analytic solution to the minimization of (\ref{rescaled}) can be obtained in the limit $z_w \gg 1$. For this, we can first minimize (\ref{rescaled}) with respect to $z_w$ by keeping only the dominant terms in a large $z_w$ expansion
\beq
z_w = 64  \pi^3 \left( \frac{ q\coth{(2 b \pi)}}{b} \right)^{3/2} \qquad {\rm for} \qquad z_w \gg 1 \ .
\eeq
We then have to minimize
\bea
\frac{1}{ \Lambda^2 }{\cal E}(q;b)=
\frac{64 \pi ^4 q^2 \coth{(2 b \pi)}}{b}+ 16 b^{3/2} \pi  \left(q^2 (\coth{(2 b \pi)})^2\right)^{1/4}
\eea
with respect to $b$. By considering only the dominant terms in the large $q$ limit, we obtain the result
\beq
\label{sollargeq}
b= \frac{2 \, 2^{1/5} \pi^{6/5} q^{3/5}}{3^{2/5}}  \qquad {\rm and} \qquad z_w= 16 \, 2^{1/5}3^{3/5}  \pi^{6/5} q^{3/5} \qquad {\rm for} \qquad q \gg 1 \ ,
\eeq
and the energy density is
\bea
\label{largeQsolution}
{\cal E}(Q)= 80 \left(\frac{2^4 \pi^{14} Q^{7}}{3^{3} \Lambda^{4}}\right)^{1/5} \qquad  {\rm for} \qquad Q \gg \Lambda^2 \ .
\eea
Note that the derivative with respect to $\Delta$ is positive:
\beq
\left.\frac{\partial
{\cal E}}{\partial \Delta}\right|_{\Delta =0} \simeq   + \frac{z_w^2}{\pi^2} -  \frac{256 \pi^5 Q }{\Lambda^2}  >0  \qquad {\rm for } \qquad Q \gg \Lambda^2 \ .
\eeq
This limit $Q \gg \Lambda^2$ is exactly the opposite of the BPS limit previously considered.
We can compute the chemical potential in this limit with the thermodynamic relation
\beq
\label{muQlarge}
\mu_Q = \frac{\partial {\cal E}(Q)}{\partial Q} =  \frac{112\, 2^{4/5} \pi^{14/5} Q^{2/5}}{3^{3/5} \Lambda^{4/5}} \ . 
\eeq
The asymptotic value $\hat{A}_0$ is obtained by integrating (\ref{a0ingen}) and (\ref{a0outgen})
\beq
\hat{A}_0(\infty) = 16 \pi^2 Q \left(  \frac{1}{4B} + \frac{1}{z_w} \right)=
\frac{7  \pi^{4/5}\Lambda^{1/5} Q^{2/5}}{  2^{1/5} 3^{3/5} } 
\eeq
and this agrees with (\ref{muQlarge}) by using (\ref{muQsecondway}).

\begin{figure}[h!t]
\epsfxsize=9.0cm
\centerline{\epsfbox{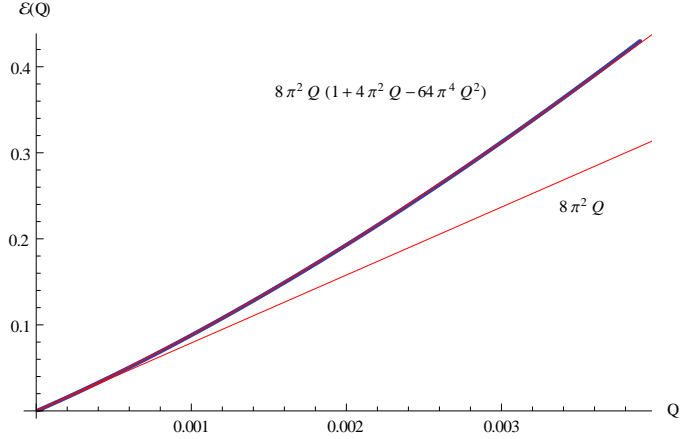}}
\caption{{\footnotesize Numerical plot of ${\cal E}(Q)$ for small values of  $Q$ obtained from the minimization of (\ref{rescaled}) for $\Lambda = 2$. This confirms the expectation from the BPS approximation (\ref{bpsapproximation}). }}
\label{lin}
\end{figure}

\begin{figure}[h!t]
\epsfxsize=9.0cm
\centerline{\epsfbox{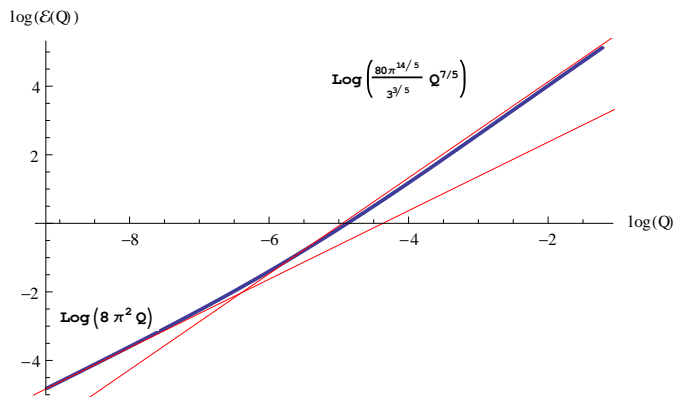}}
\caption{{\footnotesize Numerical solution of ${\cal E}(Q)$ in a log-log plot for $\Lambda = 2$. This shows the interpolation between the almost BPS limit for small $Q$ and the asymptotic solution (\ref{largeQsolution}) for large $Q$.  }}
\label{loglog}
\end{figure}

The minimization can be performed numerically for any value of $Q$. The results are shown in Figures \ref{lin} and \ref{loglog} and are confronted both with the small $Q$ and large $Q$ approximations.

We can  consider a local inertial frame  by zooming in closer to the monopole wall. In the new coordinates
\beq
\x_{\mu} =H(z_w)^{1/2} x_{\mu}  \qquad  \z =\frac{z-z_w}{H(z_w)^{1/2}} \ ,
\eeq
the metric is locally Minkowski
\beq
\label{localMinkowski}
ds^2 \simeq  d\x_{\mu} d\x^{\mu} + d\z^2 \ ,
\eeq
and this is valid as long as $|\z|$ is much smaller than the metric curvature radius, which, for $z_w \gg 1$, is the condition
\beq
|\z| \ll \frac{1}{|R|^{1/2}} \simeq \frac{3 z_w^{1/3}}{2 \sqrt{13}} \ .
\eeq
In this frame, the fields close to the wall are
\bea
F_{\z \tilde{3}} &=&
F_{z 3} = \frac{8 \pi^2 Q}{\Lambda z_w^2} = \frac{\pi^{28/5} \Lambda^{7/5}}{3^{6/5}8 Q^{1/5}}\nonumber \\
F_{\tilde{1} \tilde{2}} &=& \frac{B}{H(z_w)}   = \frac{\pi^{18/5} \Lambda^{7/5}}{3^{6/5}8 Q^{1/5}} \ .
\eea
The monopole wall lattice period $\tilde{l}$ and the transverse thickness $\tilde{\delta}$ are  given by
\beq
\tilde{l}= \frac{1}{\sqrt{F_{\tilde{1} \tilde{2}}}} \qquad \qquad
\tilde{\delta}= \frac{1}{\sqrt{F_{\z \tilde{3}}}}
\eeq
and are always smaller than the curvature radius. This justifies the approximation made before of neglecting the microscopic structure of the walls.

\section{Embedding in string theory}
\label{string}

Now we move to full string version of Sakai-Sugimoto \cite{Sakai:2004cn}.
First we briefly review the brane construction and the near-horizon geometry.
The SS model starts with the  intersection of D4, D8 and anti-D8 branes in type IIA string theory.
The gauge theory with group $SU(N_c)$ is defined on the D4-branes, which are extended along the directions $x_{0,1,2,3,5}$, and the $N_f$ flavours are provided by the D8-branes and anti-D8-branes extended along the directions $x_{0,1,2,3,5,6,7,8,9}$.  This brane intersection leaves only massless chiral fermion with one gauge and one flavor index in the low-energy theory.  Moreover the direction $x_5$ is compactified with anti-periodic boundary conditions for fermions on a circle with radius $R_5$ \cite{witten}.  With this compactification,  the low-energy of the D4 is exactly that of QCD in $3+1$ dimension with $N_f$ massless fermions.

\begin{figure}[h!t]
\epsfxsize=12.0cm
\centerline{\epsfbox{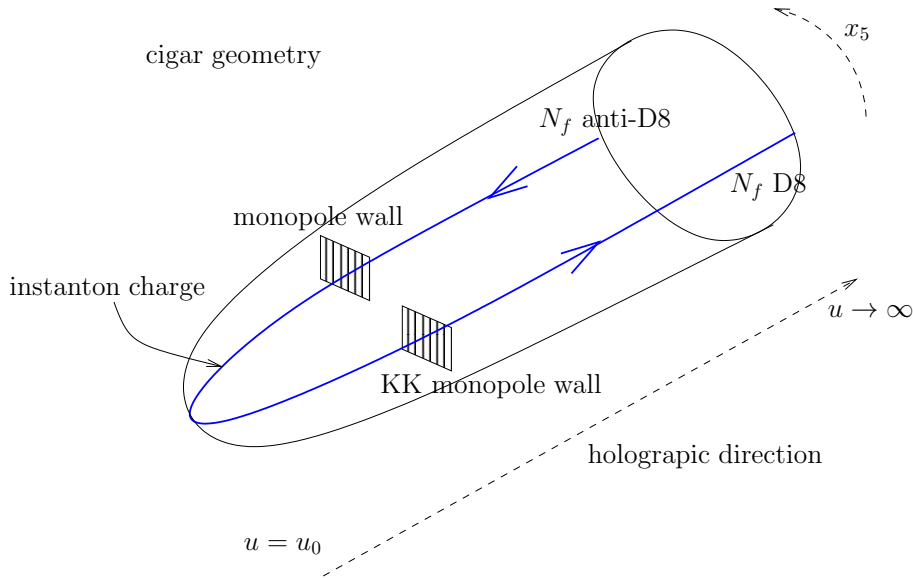}}
\caption{{\footnotesize Geometric realization of confinement and chiral symmetry breaking  in the Sakai-Sugimoto model and the embedded instanton bag on the $D8$'s worldvolume.}}
\label{cigarss}
\end{figure}

The next step is to take large $N_c$ and large $\lambda$ limits and go to near-horizon geometry of the D4-branes. The D4-branes disappear and leave a curved $AdS_5$ geometry  compactified on $x_5$ plus an $S^4$ sphere with $N_c$ units of RR flux. The D8-branes remains as physical branes in this background.
The geometry, the field strength and the dilaton are given by
\bea
ds^2 &=& \left(\frac{u}{L}\right)^{3/2} \left(dx_{\mu} dx^{\mu} +h(u) dx_5^2\right) +\left(\frac{L}{u}\right)^{3/2} \left(\frac{du^2}{h(u)}+ u^2 d\Omega_4^2 \right) \nonumber \\
F_4 &=& \frac{(2\pi)^3 \alpha'^{3/2} N_c}{V_4} {\rm vol}(S^4)\qquad  \qquad e^{\phi} = g_s \left(\frac{u}{L}\right)^{3/4}
\eea
with
\beq
h(u) = 1-\left(\frac{u_0}{u}\right)^{3}
\eeq
This is a cigar-like topology in the two-dimensional subspace $x_5, u$ 
(see Figure \ref{cigarss}).
It is like a Euclidean Schwarzschild black hole,  with $x_5$ playing the role of the Euclidean time.
The holographic direction is $u$, and  both $u \to \infty$ correspond to the UV limit of the boundary theory.
The cigar topology implies that  the two stacks of D8 and anti-D8 branes  are continuously joined together at the tip of the cigar; this is the geometric realization of chiral symmetry breaking.

We now recall  the relation between bulk and boundary theories.
The dual gauge theory is defined by  the gauge couplings $g_{YM 4}$ and $g_{YM 4}$, the number of colors $N_c$, the compactification radius $R_5$, and the 't Hooft coupling $\lambda$ related by the following:
\beq
g_{YM 4}^2 = \frac{g_{YM  5}^2}{2 \pi R_5} \qquad \quad \lambda = g_{YM 4}^2 N_c \ .
\eeq
The dynamical scale is given by
\beq
\Lambda_{QCD} = \frac{1}{R_5} e^{-cost/\lambda}
\eeq
which means that the low-energy  QCD is hierarchically separated from the Kaluza-Klein modes only for $\lambda \ll 1$. This is one limitation of the $\lambda \to \infty$ limit in the SS model.
Thus the bulk string theory has parameters $g_s$, $\alpha'$, $ L$, $u_0$, and $R_5$,  which is shared. The absence of any conical singularity at the cigar tip gives the relation
\beq
\frac{1}{R_5^2} = \frac{9 u_0}{4 L^3} \ .
\eeq
We then have three parameters to be matched between bulk and boundary. The dictionary is given by  the following three relations
\beq
\qquad
\frac{L^3}{\alpha'} = \frac{\lambda  R_5}{2}
\qquad \quad
\frac{u_0}{\alpha'} = \frac{2 \lambda}{9 R_5}
\qquad \quad
g_s \sqrt{\alpha'} = \frac{\lambda R_5}{2 \pi N_c } \ .
\eeq
Note that only three out of the four parameters, $L,u_0,g_s,\alpha'$, are need to define the boundary parameters $N_c, \lambda,R_5$.

To maintain string theory at weak coupling, we need the curvature of the bulk to be small with respect to the string scale and thus the 't Hooft coupling to be large
\beq
\frac{\sqrt[4]{L^3 u_0}}{\sqrt{\alpha'}} \gg 1 \qquad \Longrightarrow \qquad \sqrt{\lambda} \gg 1
\eeq
Moreover the string coupling has to be small
\beq
g_s\left(\frac{u_0}{L}\right)^{3/4} \gg 1 \qquad \Longrightarrow \qquad \sqrt{\lambda^{3/2}/N_c} \ll 1 \ .
\eeq
There is always a scale  $u_{UV}$ above which string theory is no longer weakly coupled. For this, we need a further restriction on $N_c$ and $\lambda$:
\beq
\frac{u_{UV}}{u_0} \simeq \frac{N^{4/3}}{\lambda^2}  \gg  1 \ .
\eeq

The low-energy theory on the D8-branes is the non-Abelian DBI action plus a  WZ  term.
Since the D8 and anti-D8 are continuously connected, we can write an action on a single extended stack of $N_f$ D8-branes.
We will mostly focus on the $N_f =2$ case.
The DBI action, in the weak field limit is just the Yang-Mills action, so that:
\bea
\label{actioncurvedcs}
S &=&     - k  \int d^4x du \  \frac{u^{5/2}}{h(u)^{1/2}}\  \frac{1}{2}{\rm tr}\left( \left(\frac{L}{u}\right)^3 {\cal F}_{\mu\nu} {\cal F}^{\mu\nu} + 2 h(u) {\cal F}_{\mu u} {\cal F}^{\mu u} \right)  + {\cal O}\left( {\cal{F}}^4 \right) \nonumber \\
&& + \frac{N_c}{24 \pi^2} \int \omega_5^{N_f} ({\cal A}) \ ,
\eea
where we raised indices with $\eta^{\Gamma\Delta}$, and the coefficient $k$ is given by
\beq
k =  \frac{L^{3/2}}{2^4 3 \pi^4 g_s \alpha'^{5/2}} \ .
\eeq
This can be written as a theory living on an effective $5$D effective metric which takes into account the effect of the dilaton:
\beq
ds = \frac{u^2}{u_0^2}  dx^2 + \frac{ L^3}{u_0^2 u h(u) }du^2 \ .
\eeq
We can  change variables from $u$ to $z$ with
\beq
1+\frac{z^2}{z_0^2}= \frac{u^3}{u_0^3} \qquad {\rm with} \qquad
z_0^2 = \frac{4 L^3}{9 u_0} = R_5^2 \ .
\eeq
This brings the metric to the form (\ref{metric}), and the action (\ref{actioncurvedcs}) to (\ref{actionymcs}) in the $x^{\mu},z$ coordinates in units $z_0=R_5=1$.

Figure \ref{cigarss} how the instanton bag solution is embedded in the D8's world-volume.
The terms (\ref{actioncurvedcs}) are a truncation of the DBI action to the first YM term. The original DBI action for the D8 brane is
\beq
S_{DBI} = T_8 \int d^9x e^{-\phi} {\rm tr} \sqrt{- {\rm det} (g_{\Gamma \Delta} + 2 \pi \alpha' {\cal F}_{\Gamma \Delta})} \ .
\eeq
To check if the truncation of the DBI action to the YM term is a good approximation, we have to check if the three corrections
\beq
\label{DBIcondition}
\frac{(\alpha'B)^2 }{g_{11} g_{22}} \ , \qquad \qquad \frac{(\alpha'F_{3u})^2}{g_{33} g_{uu}} \ ,  \qquad \qquad \frac{(\alpha'\hat{F}_{u0})^2}{g_{00} g_{uu}}
\eeq
are negligible for any $u$.
In the almost-BPS limit, that is $Q \ll \Lambda^2$, the first two in (\ref{DBIcondition}) are both of order one. So, as it happens for an instanton in isolation, the YM term receives order one  corrections from the higher derivative terms.  This is mitigated by the fact that the self-dual solutions are also solutions of the DBI action in flat space \cite{Kaplunovsky:2010eh}.  Even in the large $Q \ll \Lambda^2$ limit, computed on the  solution (\ref{sollargeq}),  these higher derivative corrections remain non-negligible.

In the effective metric (\ref{metric}) the full DBI plus CS action is
\bea
\label{actiondbics}
{\cal S} = -   \int d^4 x dz \  C \, {\rm tr} \sqrt{-{\rm det}\left(  g_{\Gamma \Delta} + \frac{2}{C^{1/2}}{\cal F}_{\Gamma \Delta}\right) }  +  \frac{8}{3 \Lambda} \int d^4 x dz \ \omega_5({\cal A}) \  \nonumber \\
\eea
where 
\beq
C 
= \frac{\Lambda^2}{16 (1+z^2)^{1/2}} \ ,
\eeq
and we  used the same rescaling as (\ref{rescalingsl}).
For the almost-BPS regime, from (\ref{bpsrelationqbzd}) and (\ref{minzwbps}), the non-Abelian field is $B= \Lambda/\sqrt{6}$ inside the instanton bag. The correction from the higher derivative terms is thus or order $B/C^{1/2}= 4/\sqrt{6}$ which is indeed non-negligible and does not depend on $\Lambda$. 
 A detailed study of the instanton bag solution using  the full DBI+CS action is beyond the scope of this paper.  We do not expect, at least in the almost-BPS regime,  a qualitative change in the solution.

\section{Chiral symmetry restoration}
\label{chisymrest}

The purpose of this section is to discuss the phenomenon of chiral symmetry restoration at high-density.

Let us first discuss the basic features of chiral symmetry breaking in the SS model with the linear expansion in eigenmodes.
The theory lives in the effective geometry (\ref{metric}), where the left and right flavour branes  correspond to the left and right limits of the holographic coordinate $z \to \pm \infty$.  The YM action  in this metric is
\beq
S = \int d^4x dz\  {\rm tr} \left( - \frac{1}{2 H(z)^{1/2}} {F_{\mu\nu}}^2 - H(z)^{3/2} {F_{\mu z}}^2  \right) \ .
\eeq
We linearly expand  around the vacuum  state
\beq
A_{\mu} =  \sum_n \, B_{\mu}^{n}(x_{\mu}) \psi_{n}(z) \qquad \qquad 
A_z =  \sum_n \, \varphi^{n}(x_{\mu}) \phi_{n}(z) \ ,
\eeq
We take $\phi_{n}$ to be  the derivative of $\psi_{n}$.
The boundary conditions correspond to the vanishing of the sources at the conformal boundaries, which is
\beq
\label{bcg}
\psi_{n}(\pm \infty)=0  \qquad {\rm for} \quad n>0\ .
\eeq
This is the condition that quantizes the mesons states. $n$ odd or even correspond respectively to even or odd states and $n=0$ is a case to be treated with special care.   $\psi_{n}$   solves the equation
\beq
\label{eqpsi}
H(z)^{1/2}\partial_z (H(z)^{3/2} \partial_z \psi_{(k)}^{\pm}(z)) + k_n^2 \psi_{(k)}^{\pm}(z) =0 \ .
\eeq
The action becomes  then
\beq
S = \sum_n \left(\psi_{n},\psi_{n}\right)  \int d^4x
\left(
- \frac{1}{4} \left( \partial_{\mu} B_{\nu}^{n}  -  \partial_{\nu} B_{\mu}^{n} \right)^2
  + \frac{k_n^2}{2} \left( B_{\mu}^{n}  - \partial_{\mu} \varphi^n\right)^2  \right) \ ,
\eeq
and the metric in $\psi(z)$ functional space is
\beq
\left(\psi_1,\psi_2\right) = \int dz H(z)^{-1/2} \psi_1(z) \psi_2(z)
\eeq
with $\psi_n$, $\psi_m$ that are orthogonal if $n \neq m$.
For $n\geq 1$ we can eliminate $\varphi$'s with a gauge transformation, and  the action is then that of a  vector boson with mass $k_n$.  When $n=0$ also $k_0 = 0$, and this gives the action of the pion.  This is a special case because the boundary condition (\ref{bcg}) is not satisfied but nevertheless there is no source at the boundary.

If we expand around another state,  we need the particular multi-instanton background which solves the finite density problem.  The spectrum in the dual theory always consists in a tower of vector bosons plus the pions.

Vector mesons come in pairs, one   vectorial $V$  and one  axial $A$, under the parity  symmetry $z \to -z$.
In our conventions, these correspond respectively to the choice of $n$ being odd or even.
One of the simplest observables that probes chiral symmetry breaking is the mass splitting  between the axial and vectorial states:
\beq
\eta_m = \frac{M_{2m} - M_{2m-1}}{M_{2m} + M_{2m-1}}
\eeq
This test can be performed at every level $m>0$, although the lower ones give, in general, the biggest $\eta$.
If $\eta_m$ is different from zero,  chiral symmetry is broken.
If $\eta_m$ is zero, or `almost' zero, the V and A states are degenerate, and chiral symmetry is restored, or `almost' restored.
In the vacuum,  the two left and right branes  are connected by the geometry, and chiral symmetry is thus broken.
This may not be the case in the presence of something in the middle that could prevent communication between the two sides.

Chiral symmetry is  restored, for example,  at  high temperature in the SS model.
Introducing finite temperature leads to  competition between the Euclidean time circle $\tau$  and the $x_5$ circle on which to close the topology.  This may lead to a phase transition with chiral symmetry restored.
In this case, the geometry of space-time has an Hawking-Page phase transition, and the two branes  become  disconnected by the presence of a horizon.
This is a drastic change in the topology, and the $V$ and $A$ states become absolutely  degenerate, i.e. $\eta_m$ is exactly zero for every level $m$.
At finite chemical potential, the situation is more subtle.  We will be mainly interested with zero temperature case,  so the topology of space-time remain unchanged (\ref{metric}).
Chiral symmetry restoration can be explained just within the effective action (\ref{actionymcs}).

We consider first a toy model that illustrates a simplified version of the phenomenon we want to discuss. 
This is the  quantum mechanical problem of a particle in  a double-well potential with Hamiltonian
\beq
H  = - \frac{1}{2} \left(\frac{d}{dx}\right)^2 + \frac{v^2}{2} (x^2 - 1)^2
\eeq
The potential has a parity symmetry $x \to -x$.
In case of a very large potential barrier, $v \gg 1$, we can approximate the eigenstates energies as
\beq
E_{m , \pm}  \simeq \frac12 + m  \mp {\cal O} \left(e^{-4v/3}\right) \qquad \qquad {\rm for} \qquad
m \ll  v
\eeq
with $m \in \mathbb{N}$.
These states are localized near the two vacua of the potential and come in pairs.
The potential provides a barrier for the states that have energy below its maximum height. For the states below this barrier, we have an approximate degeneracy between V and A states. The splitting between odd and even states can occur only through the tunneling below the barrier, and it is thus suppressed exponentially by the instanton action $e^{-4v/3}$.
The situation  in  HQCD at high-density, as we are going to see,  is  analogue to this toy model. The instanton charge provides a potential barrier between the left and right boundaries.
We then have to determine under which conditions the barrier can  seal the two sides from communication with each other, at least for some of the low-energy states, thus providing a mechanism for an effective chiral symmetry restoration.
Note that, as for the toy model, the chiral symmetry restoration can  never be exact but only up to exponentially small terms.   This is expected because, unlike the high temperature case,  the left and right branes are always connected by the geometry.

\begin{figure}[h!t]
\epsfxsize=8.0cm
\centerline{\epsfbox{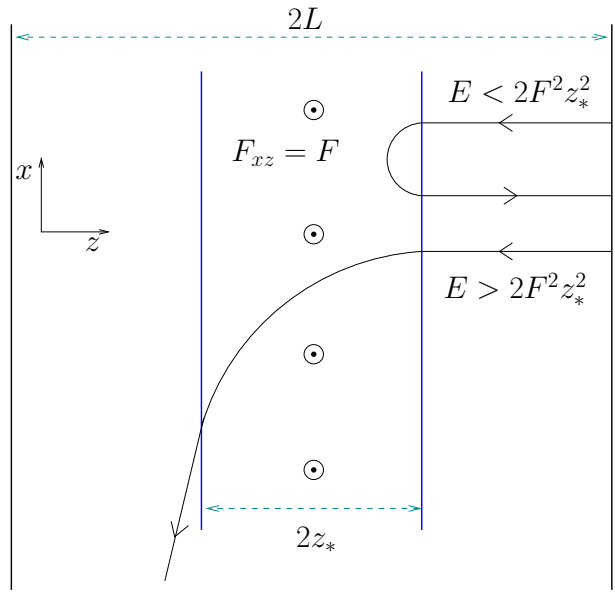}}
\caption{{\footnotesize  Charged particle tunneling a  magnetic field strip.}}
\label{tunneling}
\end{figure}
We now consider a second toy model, which is a step closer to the real thing.
We take a charged particle  living on a strip $-L<z<L$ and $-\infty<x<\infty$,  and coupled to a background gauge field. The equation of motion  is the following covariant version of the Klein-Gordon equation
\beq
\left(\frac{1}{2}(\partial_t - i A_t)^2 -\frac{1}{2}(\partial_x - i A_x)^2  -\frac{1}{2}(\partial_z -i A_z)^2 \right) \psi(t,x,z) = 0 \ .
\eeq
In a static background we can solve the eigenvalues equation
\beq
\label{operatoranalogy}
\left( -\frac{1}{2}(\partial_x - i A_x)^2  -\frac{1}{2}(\partial_z -i A_z)^2 \right) \psi_n(x,z) =  \epsilon_n \psi_n(x,z) \ .
\eeq
This is the same of the Schrodinger equation for of a non-relativistic quantum particle with  mass$/\hbar^2 =1$.
This analogy may  be useful.
For states which do not have momentum in the $x$ direction, the eigenvalues $\lambda_n$ are related to the mass in the holographic interpretation, precisely
\beq
\epsilon_n = \frac{M_n^2}{2} \qquad {\rm for} \quad k_x=0
\eeq 
In the vacuum $A_{x,z}=0$, the spectrum of the operator (\ref{operatoranalogy}) is $\epsilon_{n} = (1+n)^2/L$, where $n$ being odd or even corresponds to the parity with respect to $z \to -z$ and  clearly there is no degeneracy between V and A states.
We then turn on  a constant magnetic field $F_{x z}=F$  in a smaller strip $-z_*<z<z_*$ with $z_*<L$.
The phenomenon of magnetic trapping is quite clear by considering the classical particle trajectories in Figure \ref{tunneling}. This in the sense of the non-relativistic analogy mentioned before. For a particle to be able to cross the magnetic strip, the radius of the trajectory in the constant magnetic field zone $\sqrt{2 \epsilon}/F$ must be at least equal to the strip size $2z_*$.
There is an effective energy barrier $\simeq (F z_*)^2$ between the left and right regions.
The left and right states, if confined in the regions without the magnetic field, have energies
\beq
\label{assumesealing}
\epsilon_{m; \, V,A} =\frac{ \pi^2 (1+m)^2}{2(L-z_*)^2} \mp {\cal O} \left(e^{   - \propto \, F z_*^2}\right)
\eeq
with $m \in \mathbb{N}$:
\beq
\label{penetrationlength}
l_p \simeq \frac{1+m}{(L-z_*) F} \ .
\eeq
The condition to have V and A degeneracy is  $l_p \ll z_*$,  which,  for the lowest state $m=0$, is
\beq
\label{condpass}
 \frac{1}{(L-z_*) F } \ll z_* \ .
\eeq

To be more explicit, we take the following gauge to reproduce the desired magnetic field
\bea
&A_x = 0 \qquad \qquad \quad & z_* \geq z \geq L \nonumber \\
&A_x = F (z-z_*) \qquad \qquad \quad &-z_* \leq z \leq z_*  \nonumber \\
&A_x = - 2 F z_* \qquad \qquad \quad  &-L \geq z\leq - z_* 
\eeq
with $A_t = A_z =0$. The eigenstates can be written as
\beq
\label{sep}
\psi(t,x,z) = e^{i k_t t -i k_x x } \psi_n(z) \ ,
\eeq
where $k_t^2 -k_x^2 = M_n^2$ is the mass square from the $x,t$ perspective.
With this  ansatz, the eigenvalue equation becomes
\beq
-\frac{1}{2}\psi_n(z)'' + \frac{(F(z-z_*)-k_x)^2}{2} \psi_n(z) = \frac{M_n^2}{2} \psi_n(z) \ ,
\eeq
which is the analogue of (\ref{eqpsi}).  The wave function $\psi_n(z)$ is exponentially dumped in the magnetic field region. This damping is smaller if we increase the particle mass $M_n$ and/or if we give a momentum in the $x$ direction, as is also clear from Figure \ref{tunneling}. For the low-energy states, the penetration length is given by formula (\ref{penetrationlength}).

\begin{figure}[h!t]
\epsfxsize=8.0cm
\centerline{\epsfbox{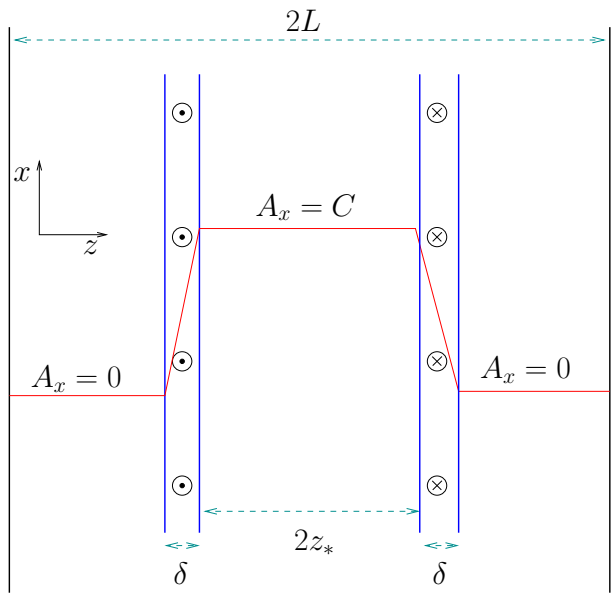}}
\caption{{\footnotesize  Charged particle tunneling a `pure gauge' strip.}}
\label{tunnelingothereffect}
\end{figure}
We now consider a third toy model (Figure \ref{tunnelingothereffect}), which contains yet a different effect that will have to be considered when we will deal with the real thing.
We take the same  charged particle as before living on a strip $-L<z<L$ and $-\infty<x<\infty$,  and coupled to a background gauge field. 
We then turn on  a pure gauge field field $A_{x}=C$ with $C$ a constant in a smaller strip $-z_*<z<z_*$ with $z_*<L$:
\bea
&A_x = 0 \qquad \qquad \quad &z_* + \delta \geq z \geq L \nonumber \\
&A_x = C  \qquad \qquad \quad  &-z_* \leq z \leq z_*  \nonumber \\
&A_x = 0 \qquad \qquad \quad    &-L \geq z\leq - z_*-\delta 
\eeq
with $A_t = A_z =0$. The two walls at the edges of the strip have thickness $\delta \ll z_*$. Here a magnetic field $F_{xz}$ is inevitable turned on to have the matching of the gauge field. We assume that this happens in the simplest way, a linear function homogeneous in $x$. 
So the magnetic field inside the two walls is respectively $F_{xz} = \pm C/\delta$.

There is now a massless eigenstate that can propagate inside the strip since it is a pure gauge. The only difference is that this state must have a phase changing in the $x$ direction to cancel the constant field $e^{i C x}$. So the main source of tunneling between the left and right side, takes place inside the two walls of thickness $\delta$, and is just given by the overlap between the massless states inside and outside the strip. The penetration length now given by
\beq
l_p \simeq \frac{(1+m) \delta}{(L-z_*)C}
\eeq
The condition to have V and A degeneracy is  $l_p \ll \delta $, that,  for the lowest state $m=0$, is
\beq
\label{condpassdue}
 \frac{1}{(L-z_*) C } \ll 1 \ .
\eeq
Note that $\delta$ disappears from this condition.

Now we finally consider the real problem of HQCD in the instanton bag background. The problem is quite complex, and we cannot provide an analytic solution for the wave functions and the spectra as in the the vacuum state (\ref{eqpsi}). We have to rely on  analogies with the previous toy models. There are two different sources of tunneling we need to consider, one is mimicked by second toy model of Figure \ref{tunneling}, the other by the third toy model of Figure \ref{tunnelingothereffect}.

The theory is defined on the metric (\ref{metric}), which is effectively a box with finite size, and this is what provides the quantization of the vector mesons' masses. 
The analogy with the previous two toy models is that the size of the strip $L$ is the curvature scale.

We begin from the almost-BPS limit, which is valid in the region of densities $\Lambda^{3/2} \ll Q \ll \Lambda^2$. The $\Lambda^{3/2}$ lower bound is when the instantons begin to populate the holographic direction, and it also coincides with the microscopic wall structure, $l \simeq \delta \simeq 1/\sqrt{\Lambda}$,  being smaller than the distance between the two walls, which is of order $Q/\Lambda^2$.  The upper bound coincides with the wall position being much smaller than the curvature scale $z_w \simeq Q/\Lambda^2 \ll 1$. 
In this case, the wave length of the  particle confined in the  empty sides is of order one (this is the analogue of $1/(L-z_*)$ in the toy model).

The first effect to consider is the tunneling as in Figure \ref{tunneling}.
Between the two walls the fields are 
\beq
\label{oscillations}
F_{12} = F_{3z} = B\, 
U(x_1,x_2)  t_{{\rm su}(2) } U(x_1,x_2)^{-1} 
\eeq
A gauge transformation $U(x_1,x_2)$ is necessary, as we discussed in Section \ref{instbag},  for the gauge fields not to have Dirac string singularities at the both boundaries.
The magnetic fields oscillate in all the su$(2)$ generators.
The vector boson state are instead waves coming from the empty sides in a fixed generator of the su$(2)$ algebra.

 The $F$ field of the toy model in Figure \ref{tunneling} is the analogue of this $F_{z x_3}$ field. We will neglect the  magnetic field $F_{12}$  for simplicity.
First, we  check if the  oscillations of the magnetic field generator, are fast enough respect to be momentum of the wave hitting the monopole wall from the empty side. This is indeed the case because the momentum of the waves confined in the empty regions is or order one while the momentum of the monopole wall lattice  is $1/l \simeq  1/\sqrt{\Lambda}$. The the vector bosons states do not have enough energy to resolve the microscopic structure of the wall, neither to see the fluctuations of the magnetic fields (\ref{oscillations}).
Then we have to check if the size of the magnetic strip is large enough to separate the left and right sides.
 The penetration length is of order $1/F\simeq 1/\Lambda$ and is much smaller than the wall's distance $z_w \simeq Q/\Lambda^2$ in this regime.  Therefore, the two sides are indeed separated by a potential barrier, at least for the channel described by Figure \ref{tunneling}.

There is yet another possible source of tunneling. The monopole wall solution is essentially an Abelian solution far from the wall.
In the Dirac gauge, the fields are all directed in one particular direction in the algebra su$(2)$, see (\ref{brelation}), (\ref{adjoint}). So the states of the vector boson fields which are directed in the same direction are completely transparent to the magnetic fields and they pass through the region between the two walls as free fields. The hedgehog gauge does not make this massless states disappear, they just have particular winding in the $x_1$ and $x_2$ direction to compensate for the gauge transformation. This is exact analogue to what happens in the toy model of Figure \ref{tunnelingothereffect}. The condition for having a barrier is that the wave length of the particle confined in the empty sides must be much smaller that the microscopic wave length $1/l$, and this we already checked to be the case.

We then go to the high-density limit $Q \gg \Lambda^2$, and we have to compute the equivalent of the energy scale (\ref{assumesealing}) in the toy model.
For this, is convenient to go first in the $\eta$ coordinate where the metric is conformally flat, which for $z \gg 1$ and $\eta \ll 1$ is
\beq
ds^2 = z(\eta)^{4/3} \left( dx_{\mu} dx^{\mu} +   d\eta^2 \right) \qquad \qquad z(\eta) \simeq \frac{27}{\eta^3} \ .
\eeq
The typical momentum of a  particle confined in the empty region $z>z_W$, in these coordinate, is
\beq
k_{\eta} \simeq \frac{1}{\eta_w} \simeq  \frac{3}{z_w^{1/3}} \ .
\eeq
We then use the  coordinate transformations to go in the $\tilde{x}, \tilde{z}$ coordinates, the ones in which the metric is (\ref{localMinkowski}) near the wall.
The result is
\beq
\label{momentumtilde}
k_{\tilde{z}} \simeq \frac{k_{\eta}}{z_w^{2/3}} \simeq \frac{3}{z_w} \ .
\eeq
This is the momentum of the wave hitting the monopole wall in a  locally Minkowski frame.

The next step is to compare this with the lattice size of the monopole wall. The following inequalities
\beq
k_{\tilde{z}} \ll \frac{1}{\tilde{l}} \qquad \Longrightarrow \qquad  \Lambda^{1/2} \ll Q^{1/2}
\eeq
mean that the momentum $k_{\tilde{z}}$ is never large enough to see the microscopic structure of the wall so we can average out the fluctuations of the magnetic field.
Then we want to estimate the penetration length of the wave with momentum (\ref{momentumtilde}) when it enters the  magnetic field region.
The penetration length is from the equation
\beq
 l_{p} \simeq \frac{k_{\z}}{F_{\tilde{3}\z}} \propto \frac{1}{Q^{2/5} \Lambda^{1/5}} \ ,
\eeq
which is even  smaller than the wall thickness $\delta$.
Thus, the two sides are not communicating also in the $Q \gg \Lambda^2$ limit.

 Being the chiral symmetry restoration only up to exponentially small terms, the $V$-$A$ mass splitting is never exactly zero, and in particular the pion, as a massless state in the spectrum of the theory, is always present. Testing chiral symmetry restoration with the pion would be more complicate than just computing the spectrum; it would require to check the pion self-interaction or the pion vector meson interation.  The pion wave function $\phi_0(z)$ is approximately constant in the two empty sides of the instanton bag, with opposite sign, and joined by an exponentially suppressed tail in the middle. This implies that the higher derivative interactions between pions and between pion and the massive vector bosons should be  exponentially suppressed. It would be interesting to understand between this aspect in terms of the chiral condensate as in \cite{Aharony:2008an,Seki:2012tt}. 

\section{Conclusions}
\label{concl}

In the first part of the paper, we discussed a multi-instanton solution in flat space which is  periodic in three directions and finite in the fourth direction. We called this an instanton bag because, in some opportune limit, it can be described by an homogeneous distribution of self-dual fields trapped between a monopole wall and a Kaluza-Klein monopole wall.
We  embedded  the instanton bag  in the Sakai-Sugimoto model. This is the dual of a phase of high-density baryons in the QCD-like theory defined on the boundary. The parameters of the solutions have been determined by the constrained energy minimization,  have been analytically solved in two limits of intermediate and large densities, and have been confronted with the numerical solution.

A transition from a lattice of instantons and a lattice of monopoles and KK monopoles pairs has been discussed in \cite{Rho:2009ym} under the name of `dyonic salt'. Our construction is somehow an extension of this because we show, using the monopole walls,  how to extend this to higher densities where the holographic direction is also probed. The fact that at high densities the instantons start to probe the holographic direction has been discussed \cite{Kaplunovsky:2012gb,deBoer:2012ij,Kaplunovsky:2013iza} and linked to the existence of a quarkionic phase. We showed that  that a configuration similar to the  dyonic salt can  be extended to arbitrary high densities and probe the holographic direction.

We still cannot perform the minimization over all the possible multi-instanton moduli space.  The instanton bag configuration we considered in this paper, is just one particular case,  for we can compute the fields, the energy,  and then minimize the moduli.

This instanton bag phase is intrinsically non-dilute, i.e. the individual instantons components cannot be distinguished and are larger than their average separation. Previous results in the toy model \cite{Bolognesi:2013jba} and also the qualitative analysis of Section \ref{holoqcd} showed that non-dilution is an inevitable  feature of large density solutions and becomes applicable exactly when the solitons start to populate the holographic direction.

The restoration of chiral symmetry is  related to the non-dilution of this phase. The non-Abelian field strength is continuously spread in the bulk, as opposed to a dilute phase in which it is confined to the instanton cores. This considerably affects the equation of motion for the gauge fields in the bulk, and thus creates a  barrier between the left and right branes leading to an  effective chiral  symmetry restoration.

\section*{Acknowledgments} I thank D.~Tong, M.~Blake and K.~Wong for discussions and collaboration to the initial stage of this project. I thank P.~Sutcliffe for discussions, in particular during  the closely  related   projects \cite{Bolognesi:2013nja2,Bolognesi:2013jba}.  This work was partially funded by the EPSRC grant EP/K003453/1 and now by the grant `Rientro dei Cervelli, RLM' of the Italian government.

\end{document}